\documentclass[12pt]{iopart}
\usepackage[dvips]{graphicx}
\begin{document}
\jl{3}

\letter{Diffusion of a polaron in dangling bond wires on Si(001)}

\author{M Todorovic\footnote{Now at Department of Materials, University
of Oxford, Parks Road, Oxford, OX1 3PH U.K.}, 
A J Fisher\footnote{email:andrew.fisher@ucl.ac.uk} and 
D R Bowler\footnote{email:david.bowler@ucl.ac.uk}} 
\address{Department of Physics and Astronomy, University College
London, Gower Street, London, WC1E 6BT U.K.}

\date{\today}

\begin{abstract}
Injecting charge into dangling bond wires on Si(001) has been shown to
induce polarons, which are weakly coupled to the underlying bulk phonons.
We present elevated temperature tight binding molecular dynamics simulations
designed to obtain a diffusion barrier for the diffusive motion of these
polarons.  The results indicate that diffusion of the polarons would be 
observable at room temperature, and that the polarons remain localised 
even at high temperatures.
\end{abstract}

\submitted
\maketitle

\section{\label{sec:intro}Introduction}
There are many motivations to understand the transport properties of
materials in the extreme one-dimensional limit.  Some are
technological: the logical conclusion of the historic reduction in
size of electronic components would be device elements, and passive
connections between them, that are of atomic scale.  There has been
much recent interest in structures that might act as such atomic-scale
wires, or as atomic- or molecular-scale switches.  Other reasons
relate to fundamental physics: transport in one dimension is
qualitatively different from in higher dimensions, essentially because
of the much stronger coupling that occurs between the different
possible excitations.

One particularly suitable system for the study of such effects, and
also for potential applications, is the dangling bond (DB) line on the
Si(001) surface~\cite{Shen1995}.  A local, highly one-dimensional
conducting channel is formed by the selective desorption of H atoms
from the hydrogenated surface, creating locally depassivated dangling
bonds.  Charge (produced by injection or doping) may be expected to be
strongly confined to the depassivated region.  Being an almost ideal
one-dimensional conductor, this system shows features characteristic
of strongly-coupled low-dimensional systems, including a Peierls-like
Jahn-Teller distortion~\cite{Hitosugi1999} in which neighbouring
depassivated Si atoms acquire alternating `up' and `down'
displacements normal to the surface.

It is tempting to draw an analogy with conjugated polymers, where a
similar alternation occurs in the bond length~\cite{heeger88}.  In
these systems, the coupling of injected charge to the bond
alternation results in the formation of small polarons (charge
carriers self-localized in the molecule by their own induced atomic
distortions).  Indeed, we have previously shown~\cite{Bowler2001} that
analogous excitations exist in the DB wires; our tight-binding
calculations showed that a hole becomes localized on an `up' atom,
pulling it down (i.e. making it less $\mathrm{sp}^3$-like), while an
electron becomes localized on a `down' atom, pulling it up
(i.e. making it less $\mathrm{sp}^2$-like).  From this it is clear that
there are important differences from, as well as similarities to, the
conducting polymer case, because the predominant local chemistry
driving the carrier localization involves a \textit{single\/} atom.

It might be expected that the formation of polarons would have
important consequences for the charge transport properties of the
wire. It is already known that for conjugated molecular systems in the
coherent limit, the charge transport becomes dominated by polaron
tunnelling~\cite{Ness1999}.  However, experiments on surfaces are more
likely to operate in the incoherent limit; in this case, the most
relevant quantity is the diffusion constant of the polaron along the
dangling-bond chain.  In this paper we address this issue by
performing constant-NVT tight-binding molecular dynamics for diffusing
hole polarons at a range of elevated temperatures.  We choose hole,
rather than electron, polarons because we believe our tight-binding
calculation is likely to provide a better description (see
section~\ref{sec:details}).  The tight-binding approximation is
important because it allows us to simulate large enough systems, and
for long enough times, to gain useful information on the hopping.
From the observed polaron hopping rates we are able to estimate the
activation energy to hopping as only 0.06\,eV.  Our results show that
the polaron is relatively mobile along the line, but confined
perpendicular to it even at elevated temperatures.  

The plan of the paper is therefore as follows.  First, we describe the
computational methods used.  Then, we present results for the
diffusional hopping rate, along with visualizations of the motion of
the polaron and plots of the energy of the polaron state.  We finish
with a discussion of the results.

\section{\label{sec:details}Computational Details}

The tight binding technique has been reviewed elsewhere in
detail~\cite{Goringe1997}.  We used a nearest neighbour, orthogonal
parameterisation for Si-Si bonds and Si-H bonds designed specifically
for the Si(001) surface~\cite{Bowler1998} and tested extensively in
this environment (in particular, it has been shown to reproduce
\textit{ab initio\/} simulations of the perfect, infinite dangling bond
wire rather well~\cite{Bowler2001}).  The simulations were performed
with the \textsc{Oxon} code, using exact diagonalisation.

We used periodic boundary conditions, and simulated a surface with a
slab geometry and a vacuum gap of 29\,\AA\ (the size is irrelevant in
tight binding simulations so long as it is greater than the hopping
parameters cutoff).  The unit cell was twelve dimers long and two
dimer rows wide, with one dimer row completely covered with hydrogen
and the other half-covered to yield the dangling bond wire.  There
were six layers of silicon, the bottom of which was fixed in bulk-like
positions and terminated in hydrogen, giving a total of 420 atoms.
Only the $\Gamma$-point of the reduced Brillouin zone was sampled.

The molecular dynamics simulations were performed at constant number
of particles, volume and temperature (NVT), with the temperature
maintained by a simple rescaling (if the temperature deviated by more
than 1\% the atomic velocities were rescaled to give the correct
temperature).  We used a timestep of 1\,fs, and performed simulations
for 1000 steps, giving a total time of 1\,ps.  The simulations were
carried out at temperature intervals of 100\,K beginning at 100\,K and
ending at 1000\,K.

\begin{figure}
\includegraphics[width=\columnwidth,clip]{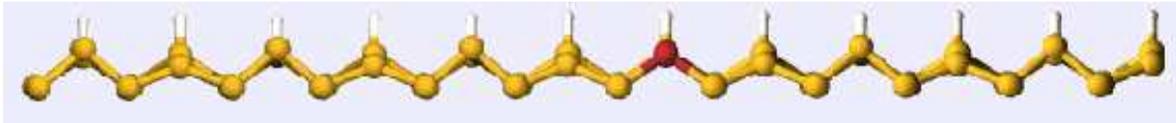}
\caption{\label{fig:holepolaron}A hole polaron (marked as a darker atom) on
a dangling bond wire on Si(001).}
\end{figure}

The starting point was a relaxed dangling bond wire with a hole
polaron already formed, illustrated in \Fref{fig:holepolaron}.  The
``perfect'' dangling bond wire consists of a series of clean Si atoms
along one dimer row on a hydrogenated surface.  As this would result
in half-filled bands, the system is unstable towards a Peierls
distortion~\cite{Peierls1955,Hitosugi1999}.  Alternate atoms in the
wire are displaced up and down, with charge transferring from the down
to the up atom.  The hole polaron is then formed by displacing a
single `up' atom of the wire downwards normal to the surface to break
the translational symmetry, and allowing the system to relax (with one
fewer electrons than required for charge neutrality).  The structure
and properties of the hole polaron (and the equivalent electron
polaron) have been discussed elsewhere~\cite{Bowler2001}.

For the MD, the atoms were given random velocities corresponding to
the appropriate Maxwell-Boltzmann distribution for the temperature
being modelled, and after equilibration, the system was allowed to
evolve for 1\,ps.  As well as monitoring the standard parameters (such
as different energies) we followed the weight of the contribution from
each atom in the system to the top-most filled eigenstate, which is
identified as the polaron state.  This gives the location of the
polaron with time, providing that it remains localised (for instance,
in the relaxed polaron, this weight is $\sim 0.4$ from the `up' atom
in the wire on which the polaron is located).  A hop was defined to
have occurred when the largest contribution changed from one atom to
another.  The occupation of the electronic states was weighted by the
appropriate Fermi function at each temperature.

\section{\label{sec:results}Results}

\begin{figure}
\includegraphics[width=\columnwidth,clip]{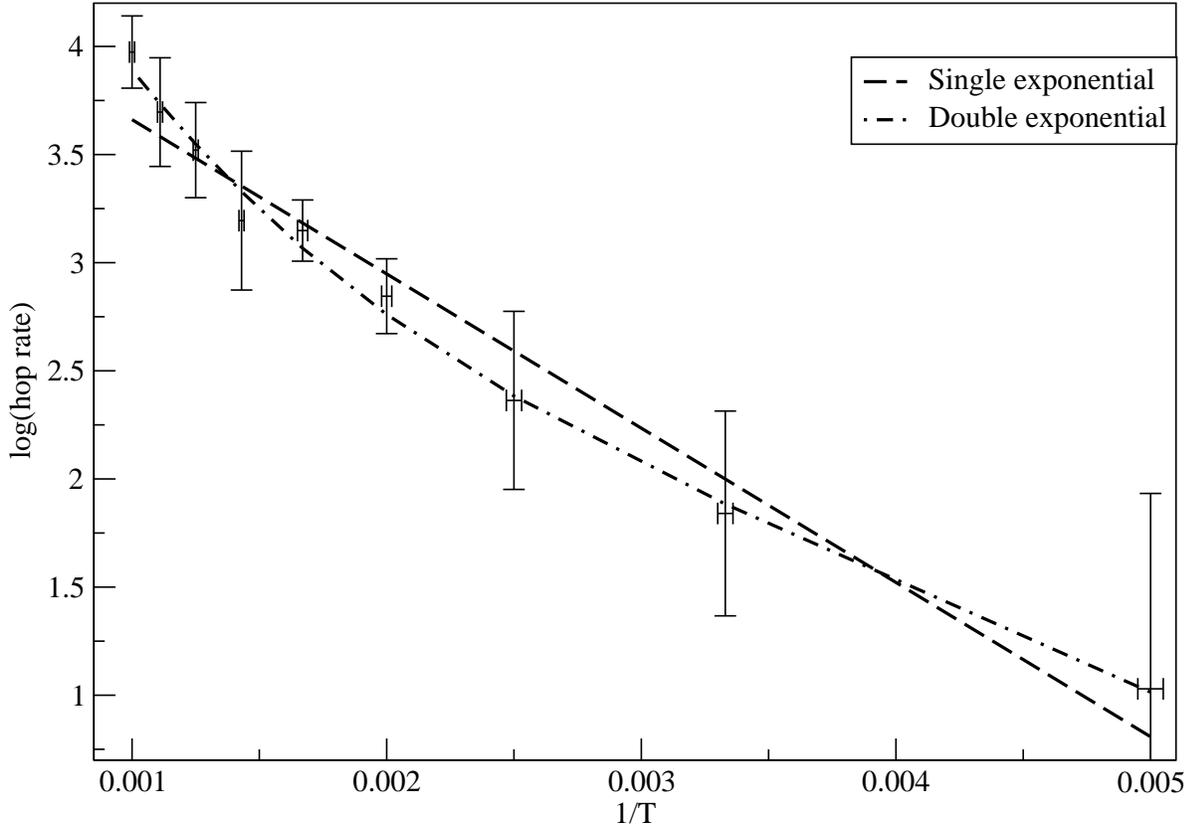}
\caption{\label{fig:hopping}The log of the rate of hops plotted
against inverse temperature.  The dashed line is a best fit to the
data for a single exponential, while the dot-dashed line is a best fit
to the data for a sum of two exponentials.}
\end{figure}

Ten MD simulations were performed for each temperature as described
above, and the number of hops at each temperature was used to give a
hopping frequency (except at 100\,K, where no hops were observed in any
of the simulations).  We assume an Arrhenius form for the hopping
rate, \textit{p\/}:
\begin{equation}
p = A \mathrm{e}^{-E_{b}/k_\mathrm{B}T},
\end{equation}
where $E_b$ is the barrier to diffusion and \textit{A\/} is the
attempt frequency \cite{Vineyard1957}. Then, by plotting the natural
logarithm of the hopping rate against inverse temperature we should
get a straight line.  The gradient gives us the barrier to diffusion
and the intercept the attempt frequency.  The graph is shown in
\fref{fig:hopping}.  The barrier is 0.06\,eV and the attempt frequency
$8\times10^{13}$ \,s$^{-1}$.

\begin{figure}
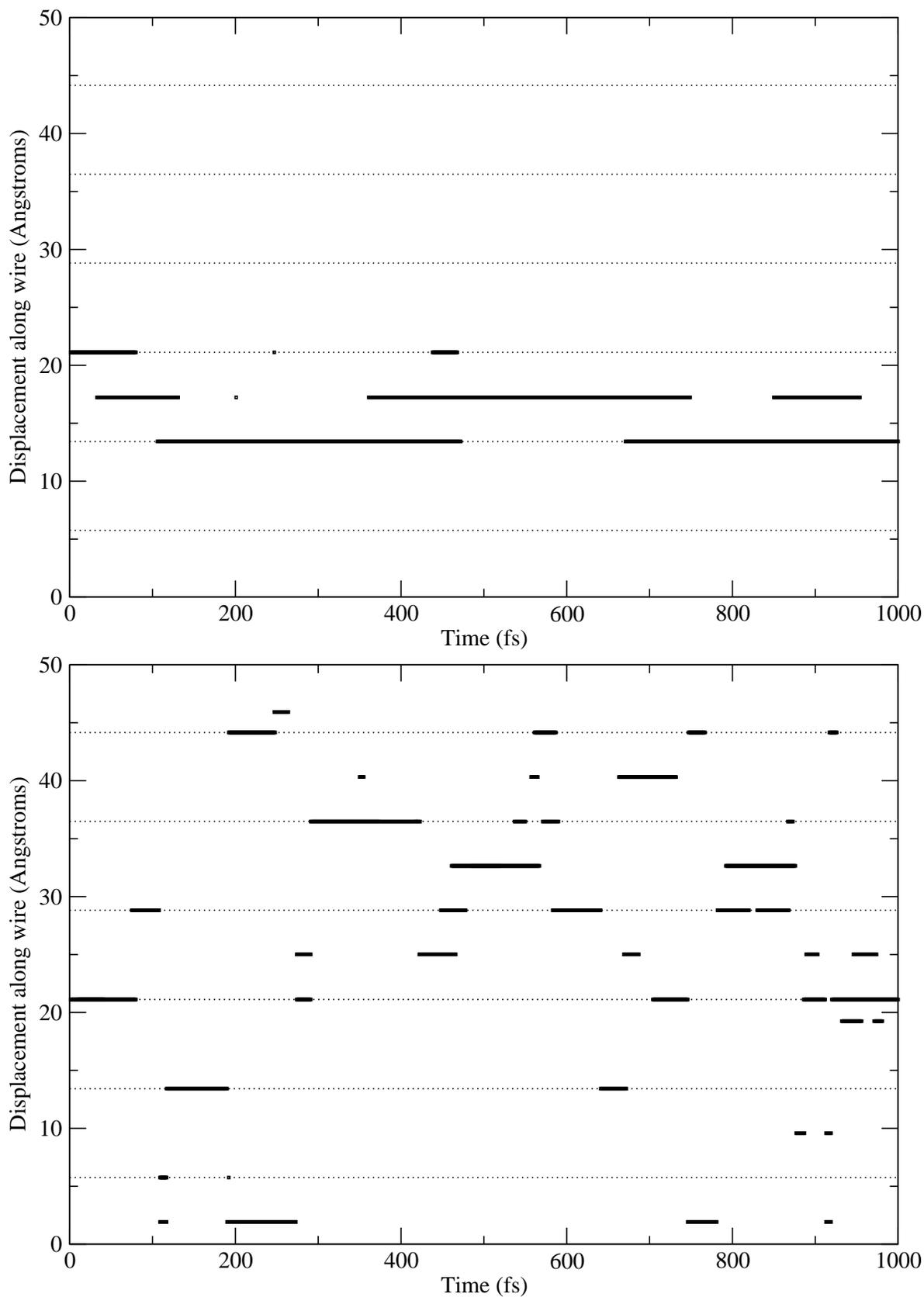

\includegraphics[width=\columnwidth,clip]{DisplacementT200}
\includegraphics[width=\columnwidth,clip]{DisplacementT700}
\caption{\label{fig:disp}Displacement of the hole polaron along the
dangling bond wire as a function of time for (a) T=200K and (b)
T=700K.  Dashed lines across the page mark the location of the ``up''
atoms in the dangling bond wire and are separated by two atomic
spacings or 7.68\AA.}
\end{figure}

On closer inspection, the graph seems to show a bend around 500\,K;
such behaviour has been seen for a small polaron
before~\cite{Norgett1973}, but our model does not contain the
necessary physics.  The bend might also be caused by one or more
alternative channels which open at high temperatures (as seen, for
instance, for H in Nb~\cite{Gillan1987,Schober1990}).  We also fitted
a sum of two exponentials, i.e.:
\begin{equation}
p = A_1 \mathrm{e}^{-E^1_{b}/k_\mathrm{B}T} + A_2 \mathrm{e}^{-E^2_{b}/k_\mathrm{B}T}.
\end{equation}

Here, we find two sets of parameters: the attempt frequency and
barrier for the first are $3\times10^{13}$ \,s$^{-1}$ and 0.043\,eV;
for the second, they are $2\times10^{14}$ \,s$^{-1}$ and 0.185\,eV.
This suggests that there may well be a number of competing channels to
the diffusion at higher temperatures.  Of course, at higher
temperature the hydrogens also become mobile~\cite{Bowler2000}, though
at these temperatures and timescales (i.e. over 1\,ps) there is no
hopping of the hydrogens.  If we were to include the quantum effects
alluded to above, this would be expected to further flatten the curve
at low temperatures.

Since we output the contributions to the topmost filled eigenstate, we
can follow the location and localisation of the polaron at different
temperatures.  Interestingly, the polaron remains localised even up to
high temperatures, and we can follow its location with time by
plotting the location of any atom on which the weight is greater than
0.1.  This is shown in \fref{fig:disp} for individual runs at two
temperatures: 200\,K and 700\,K.  The polaron moves around more at
700\,K (as would be expected) but is still localised essentially on
one atom, though there are times when it is shared between two or more
atoms.  This sharing is a prelude to a hopping attempt, and can also
be seen for the 200\,K case.

\begin{figure}
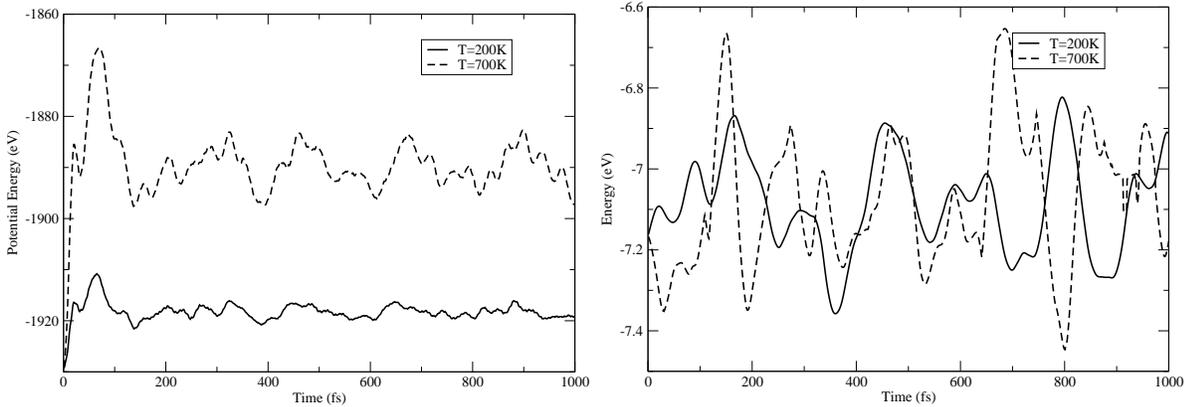

\includegraphics[width=0.495\columnwidth,clip]{PotentialEnergy}
\includegraphics[width=0.495\columnwidth,clip]{FermiLevel}
\caption{\label{fig:energy}(a) Variation of potential energy with time
for T=200\,K (solid line) and T=700\,K (dashed line) (b) Variation of
energy of the top occupied eigenstate (equivalent to the Fermi level )
with time for T=200\,K (solid line) and T=700\,K (dashed line).}
\end{figure}

The transition state for the hopping of the polaron has proved
difficult to isolate using tight binding.  We have considered two ways
of isolating it: first, mapping out a reaction pathway by constraining
an appropriate reaction coordinate at various different values and
relaxing the system at each of these points; second, analysing the MD
results to extract the atomic motion around a hop.  The first
technique failed because of the ease with which charge can be
transferred in tight binding.  We chose as our start and end points
the polaron localised on adjacent `up' atoms, and constrained the
height \textit{difference\/} between these two atoms.  Unfortunately,
as we approached a difference of zero, the charge moved around (in
particular moving to the hydrogenated ends of the dimers which were
being constrained) and the structure distorted.  We were unable to
find a transition of the polaron from one `up' atom to the next using
this technique, though we believe that it was more due to the
shortcomings of the tight binding.  The second technique failed
because the barrier is so small, and the thermal motion of the atoms
was sufficient to conceal the details of the transition.  We show this
in two ways in \fref{fig:energy}.  In \fref{fig:energy}(a), we plot
the variation of the potential energy of the system throughout the run
at 200\,K and 700\,K; this shows that, even at 200\,K, the energy was
varying by far more than the 0.06\,eV seen in the hopping of the
polaron.  In \fref{fig:energy}(b), we plot the energy of the top-most
occupied state (which is also equivalent to the Fermi energy, because
of the missing electron, and is the polaron state).  From the plot, we
can again see that the variation of the energy is larger, even for
this state, than the hopping barrier.  We have tried to correlate the
peaks and troughs seen on this plot with the hopping events, but with
limited success: there is a small correlation between the stationary
points of the energy and hopping, but not sufficient to be of
interest.  This again is due to the random thermal motion---if we were
able to follow the system for a sufficiently long time at low
temperatures (say 50\,K) we might learn more, but this would require
prohibitively long MD runs.

\section{\label{sec:conc}Discussion and Conclusions}
Our results show that, despite the self-trapping, the charge carrier
(hole in this case) is highly mobile on a timescale of 1\,ps, even at
200\,K.  Importantly, however, it remains self-trapped---and also
localized on the DB wire.  We expect that an electron polaron would
have qualitatively similar properties.  We can infer an effective
diffusion constant and mobility from our data: the diffusion constant
is $D=2Ra^2$, where $R$ is the hop rate (see \fref{fig:hopping}), and
$a$ is the hopping distance of $3.84\,$\AA.  At 300\,K, we find
$D=1.86\times10^{-6}\,\mathrm{m^2s^{-1}}$.  The mobility is found from
the Einstein relation $\mu=eD/(k_BT)$; at 300\,K we find
$\mu=7.19\times 10^{-5}\,\mathrm{m^2V^{-1}s^{-1}}$. Despite the
self-trapping, this is an appreciable mobility.

It is important to ask how much effect our adoption of the
tight-binding approximation had on these results.  As explained
previously, tight binding gives a good description of the neutral
Peierls-distorted DB wire, reproducing both the band structure and
geometry well.  These factors between them should play a dominant 
role in determining the structure and mobility of the polaron, 
implying that, despite its simplicity, the tight-binding approach
ought to describe the motion well.

In conclusion, we have shown that despite the predicted self-trapping,
hole polarons in dangling-bond wires retain considerable mobility even
at room temperature and below.  This occurs through hopping between
self-trapped configurations, with a barrier of 0.06\,eV (assuming a
single process) or 0.04\,eV (at low temperatures, assuming competing
processes).  It is possible that the mobility might be further
enhanced at low temperatures by quantum diffusion effects, not
included in our calculations.  Our calculations suggest that hopping
at low temperatures may be slow enough to be observed directly in STM,
provided that a mechanism for doping the wires can be devised.

\ack DRB acknowledges the Royal Society for funding through a
University Research Fellowship.  We are happy to acknowledge useful
discussions with Charles Bird and Robert Wolkow, Marshall Stoneham and
Jacob Gavartin.

\end{document}